# A very simple, robust and fast method for estimating and displaying average time constants of T2 decays from multiecho MRI images using color intensity projections

Keith S Cover[1]
[1]Department of Physics and Medical Technology,
VU University Medical Center, Amsterdam

Corresponding author:
Keith S Cover, PhD                                    Email: Keith@kscover.ca
Department of Physics and Medical Technology    Tel:    31 20 444-0677
VU University Medical Center                       Tel:    31 20 444-0677
Postbus 7057
1007 MB Amsterdam
The Netherlands

**Running Title:** Displaying multiecho images using color intensity projections

**Word Count:**  2,009

## Abstract

While the multiexponential nature of T2 decays measured in vivo is well known, characterizing T2 decays by a single time constant is still very useful when differentiating among structures and pathologies in MRI images. A novel, robust, fast and very simple method is presented for both estimating and displaying the average time constant for the T2 decay of each pixel from a multiecho MRI sequence. The average time constant is calculated from the average of the values measured from the T2 decay over many echoes. For a monoexponential decay, the normalized decay average varies monotonically with the time constant. Therefore, it is simple to map any normalized decay average to an average time constant. This method takes advantage of the robustness of the normalized decay average to both artifacts and multiexponential decays. Color intensity projections (CIPs) were used to display 32 echoes acquired at a 10ms spacing as a single color image. The brightness of each pixel in each color image was determined by the intensity of the corresponding pixel in the earliest image and the hue was determined by the normalized decay average. Examples demonstrate the effectiveness of using CIPs to display the results of a multiecho sequence for a healthy subject and a multiple sclerosis patient.

**INTRODUCTION**

The application of color intensity projections (CIPs) (CoverKS2006A, CoverKS2007A, CoverKS2007B) to 32 echo MRI images has yielded a novel, simple, robust and very fast method for assigning an average time constant to a T2 decay.  While the multiexponential nature of T2 decays is well known (MenonRS1992, MackayAL1994, WhittalKP1999, MooreGRW2000, Cover2006B), characterizing T2 decays by a single time constant is still widely used when differentiating between structures and pathologies (BriellmannRS2002, DardzinskiBJ1997, FerrierJC1999, GrahamSJ1996, LiessC2002, TeicherMH2000).

A simple method is presented for calculating the "normalized decay average" for any measured decay. Since it is possible to calculate a normalized decay average for any monoexponential decay, it is therefore possible to assign an "average time constant" to any decay for which the normalized decay average has been calculated. The use of the normalized decay average means that artifacts, such as even echo rephasing, tend to be averaged out. In addition, multiexponential decays also have a normalized decay average and can thus be characterized by a single average time constant.

The advantages of using the normalized decay average to determine an average time constant of a T2 decay was realized after the application of the percentage time version of color intensity projection to multi echo MRI images (CoverKS2006A). CIPs is a display technique used to summarize the changes over a series of grayscale images in a single color image. While any pixel that is constant over the series of grayscale images will appear as a grayscale pixel of the same brightness in the CIPs image, changing pixels will be represented by a variety of colors and shades. In CIPs the hue of the color is determined by the normalized average value of the pixel. One of the strengths of CIPs is that it is calculated for each pixel independent of all other pixels. Thus, any pixel to pixel variations are easily observed in a CIPs image.

The characterization of multiexponential decays by a single time constant and the display of the CIPs image for both a healthy control and multiple sclerosis (MS) patient are used to demonstrate the simplicity, speed, and robustness of the display method.

**METHODS**

To acquire the multiecho images, a GE Signa 1.5T MRI was programmed to scan a single 5mm slice with a Carr-Purcell-Meiboom-Gill (CPMG) (MeiboomS1958) sequence (WhittallKP1997, PoonCS1992, CoverKS2006B). The sequence acquired 32 echoes with a 10ms spacing followed by 16 echoes with a 100ms spacing but only the first 32 echoes were used in this paper. The images used to demonstrate the technique described in this paper were acquired from a healthy control and an MS patient. Both subjects were scanned with 128 phase and 256 frequency encodes. The field of view was 220cm by 220cm and the TR was 2.0 seconds. These images were acquired while the author was a PhD student at the University of British Columbia in a study approved by the university's ethical board.

To provide calibration information, two standards were wrapped around the head of each of the two subjects before their scans. The standards were intended to have nearly monotonic T2 decays the actual values roughly in the 20ms and 80ms. The stability of the standards over the time between the two measurements is uncertain.

The calculation of the CIPs followed the description provided in CoverKS2006A. CIPs uses the brightness, saturation and hue (BSH) scheme to quantify a color of a pixel rather than the more common red-green-blue (RGB) scheme. In BSH, the brightness determines the brightness of the pixel. The saturation has a range of 0 to 1 where 0 indicates the pixel is a shade of gray that is described completely by its brightness and 1 indicates the pixel has a bright color. The hue ranges from 0 to 1 with the colors corresponding to red-yellow-green-light blue-blue-purple-red. To avoid the confusion between the red at both 0 and 1, only hues ranging from red up to blue are used in the CIPs images.

When calculating a CIPs image, the color of each pixel is calculated independent of all other pixel values. First, for all value of each pixel over the multiecho images, the maximum, mean and minimum values are calculated. The brightness, saturation and hue of the color image are then calculated with the following formulas:

brightness = max                    [1]

saturation = (max-min)/max          [2]

hue = (mean-min)/max                [3]

Examination of these equations demonstrates that, for pixel values that remain constant over a series of images, CIPs yield pixels with zero saturation and a brightness equal to the pixel value. However, if the pixel values change over the echoes, the larger the change the more color is introduced into the pixel.

The normalized decay average is defined to be the value of the hue given by equation [3]. It ranges between 0 and 1. To convert between normalized decay averages and average time constants, a table was calculated of the normalized decay averages for a selection of monoexponential decays with a range of time constants. To match the multiecho acquisition sequence used to acquire the multiecho images, the monoexponential decays were sampled at 32 time points with a 10ms spacing.

To enhance the contrast of the CIPs images, the hue and brightness values have been windowed before display of the images. For all the CIPs images presented in this paper, the brightness values from 10% to 90% of the maximum brightness over all pixels of all images were mapped to the darkest and brightest pixels respectively. Two different windowings of the hue where applied for each of the subjects. First, the values from 0.1 to 0.4 were mapped to red through blue. Second, the values from 0.2 to 0.3 were mapped to red through blue. As would be expected, the second windowing is much more sensitive to changes in the average time constant. As is standard practice with windowing, linear interpolation was used to calculate the values between the end points. The CIPs for both subjects were calculated in exactly the same way so their CIPs images can be easily compared.

**RESULTS**

Table 1 provides the map between the average time constant and the normal decay average assuming monoexponential decays. The nature of the relationship between these two values, which is strictly monotonically increasing, is clearly evident. It is also evident that the range of normalized decay averages that are easily measured also map to the range of time constants that are typically of interest for in vivo MRIs.

Figures 1(a) and 2(a) show CIPs images of the healthy subject and the MS patient. The range of the hue is shown in the color bar in each of the two images. As described in the method, the hue spans red-yellow-green-light blue-blue with the red representing the shortest relaxation time and the blue representing the longest. From Table 1, red corresponds to 27ms, green to 61ms and blue to 111ms. Figures 1(b) through 1(e) and 2(b) through 2(e) shows a few of the 32 echo images used to generate the CIPs.

Two standards are visible in both Fig. 1 and 2. In Fig. 1 the roughly 20 ms standard that shows as red while the roughly 80ms standard shows as green. The different colors of the standards demonstrates the ease with which multi-echo CIPs images allow the estimation of relaxation time from a single image. The time constant of the cerebral spinal fluid (CSF), which shows as the blue pixels in the ventricles of Fig. 1 is about 300ms. The standards in Fig. 2 have slightly different color indicating faster relaxation times than Fig. 1. It is unclear if this difference is due to the drift of standards between scans of the two subjects. Also, there is always the possibility that imperfections in the multiecho relaxation sequence may result in different T2 time constants for the same substance located in different regions of the scanner.

Figures 3 and 4 show average time constants ranging from 61ms (red) to 111ms (blue). The yellow and green standard on the left side of Fig. 3 shows a gradual variation of the average time constant over the standard. The consistent hue of adjacent pixels demonstrates the reproducibility of the average time constant indicating a useful signal to noise ratio. However, the gradual change from red to green from the left to right of the standard indicates the average time constants are getting slower. The reason for this slowing is unknown but the CIPs image clearly demonstrates it exists.

**DISCUSSION**

The CIP of the multi-echo data allows the easy identification of various structures. Starting at the green standard in Fig 1(a) and moving towards the center of the brain shows the scalp (green), muscles above the ear (red), skull (black), a thin band of CSF (blue), gray matter (dark green), white matter (light green) and ventricles filled with CSF (dark blue). In areas adjacent to the ventricles there are a few yellow areas indicating a faster relaxation time. The vessel located in the lower part of the image is red because moving blood decays very quickly in an MRI.

Careful examination of the image also shows many areas with the dark blue hue corresponding to CSF.

Figure 2(a) shows similar structures and colors to Fig 1(a). There are two notable differences. The first is that the ventricles contain gray pixels as well as blue. The gray pixels indicate the decay of the CSF was so slow that is considered constant due to the windowing of the hue. The second notable difference is the presence of a large lesion in the white matter. The lesions generally show up as light blue, a reflection of their longer relaxation time than the white matter. The lesion has a dark blue core, indicating a particularly slow relaxation time.

Figure 3 is particularly interesting. Several white matter structures clearly stand out with the faster time constants indicated by yellow and red. Again, the well defined structures in the various hues indicate the average time constant is being calculated in a reproducible manner. In contrast, the MS patient in Fig. 4 shows few of the white matter structures visible in Fig. 3. Wide spread damage to white matter in MS patients has been reported in the literature (FilippiM1995, VrenkenH2005). Fig. 4 has been reviewed with different window levels to see if the white matter structure simply have a slower time constant than Fig. 3 but few potential white matter structures were found. Whether the lack of white matter structure is an indication of the disease, a natural variation in the general population or problem with the multiecho sequence will require additional study.

An interesting characteristic of Fig. 3 is the lack of yellow and red pixels in the lower left part of the brain as compared to the rest of the brain. As with the variation of the average time constant across one of the standards, it is unclear whether these slower average time constants are due to an imperfection in the multiecho sequence or whether is actually reflects a variation in the anatomy of the subject.

The application of CIPs to multiecho MRI sequences could be modified in several ways. While 32 evenly spaced echoes have been used in this paper, unevenly spaced echoes are sometimes acquired. For unevenly spaced echoes, calculation of the normalized decay average might be improved by weighting each echo by the time since the previous echo, or some similar weighting. In this paper the first echo of the 32 echoes was used to determined the brightness of each pixel in the CIPs. It is possible to use a later echo instead. Later echoes may be an advantage since they sometimes provide better contrast.

The application of CIPs to 32 echo relaxation sequences has yielded a surprising robust method for estimating and displaying an average time constant from T2 decays. While this initial examination of the method has yielded some very encouraging results, obviously much additional study is required to assess the value of this method.

**Acknowledgements**

This work was funded by the Department of Physics and Medical Technology at the VU University Medical Center, Amsterdam.

**Table 1**
The map between normalized decay averages, which were generated from monoexponential decays, and the time constants of the monoexponential decays. The decays were sampled at 32 points with a 10ms spacing.

| Time constant (ms) | Normalized Decay Average |
|---|---|
| 26.70 | 0.1000 |
| 43.04 | 0.1500 |
| 60.77 | 0.2000 |
| 82.22 | 0.2500 |
| 111.34 | 0.3000 |
| 156.72 | 0.3500 |
| 243.5 | 0.4000 |
| 497 | 0.4500 |

**Figure Captions**

**FIG. 1.** (a) CIPs of a healthy subject. The hue is windowed so that red, green and blue hues correspond to average time constants of 27ms, 82ms, and 244ms respectively. Images (b), (c), (d), and (e) show some of the component images (at 10, 60, 100 and 320ms) used to generate the CIPs.

**FIG. 2.** Images processed and displayed in the same way as Fig. 1 but for the MS patient.

**FIG. 3.** CIPs of the healthy subject using the same 32 echoes as Fig. 1 but with the hue windowed so that red, green and blue correspond to average time constants of 61ms, 82ms, and 111ms respectively.

**FIG. 4.** Images processed and displayed in the same way as Fig. 3 but for the MS patient.

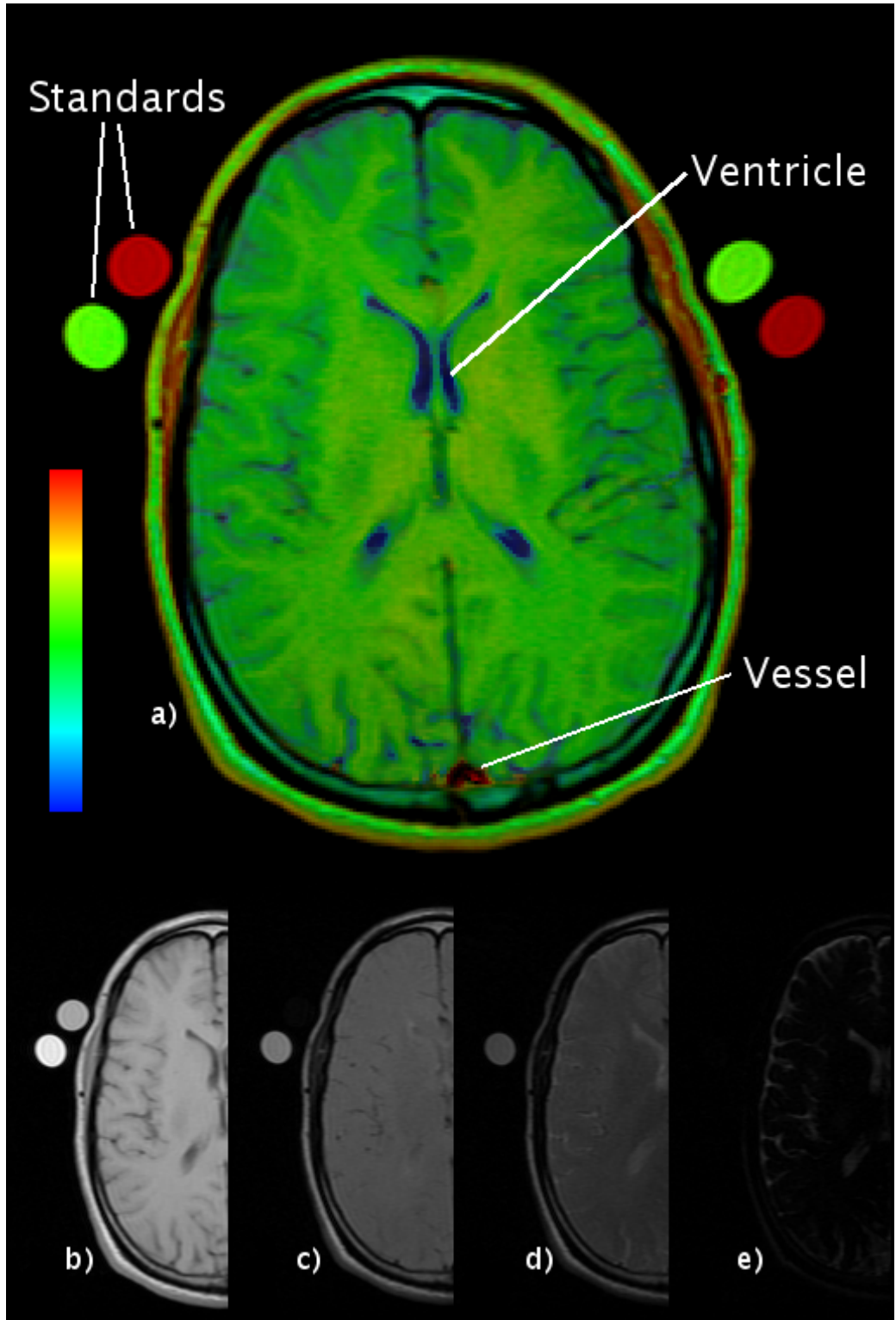

**Fig 1**

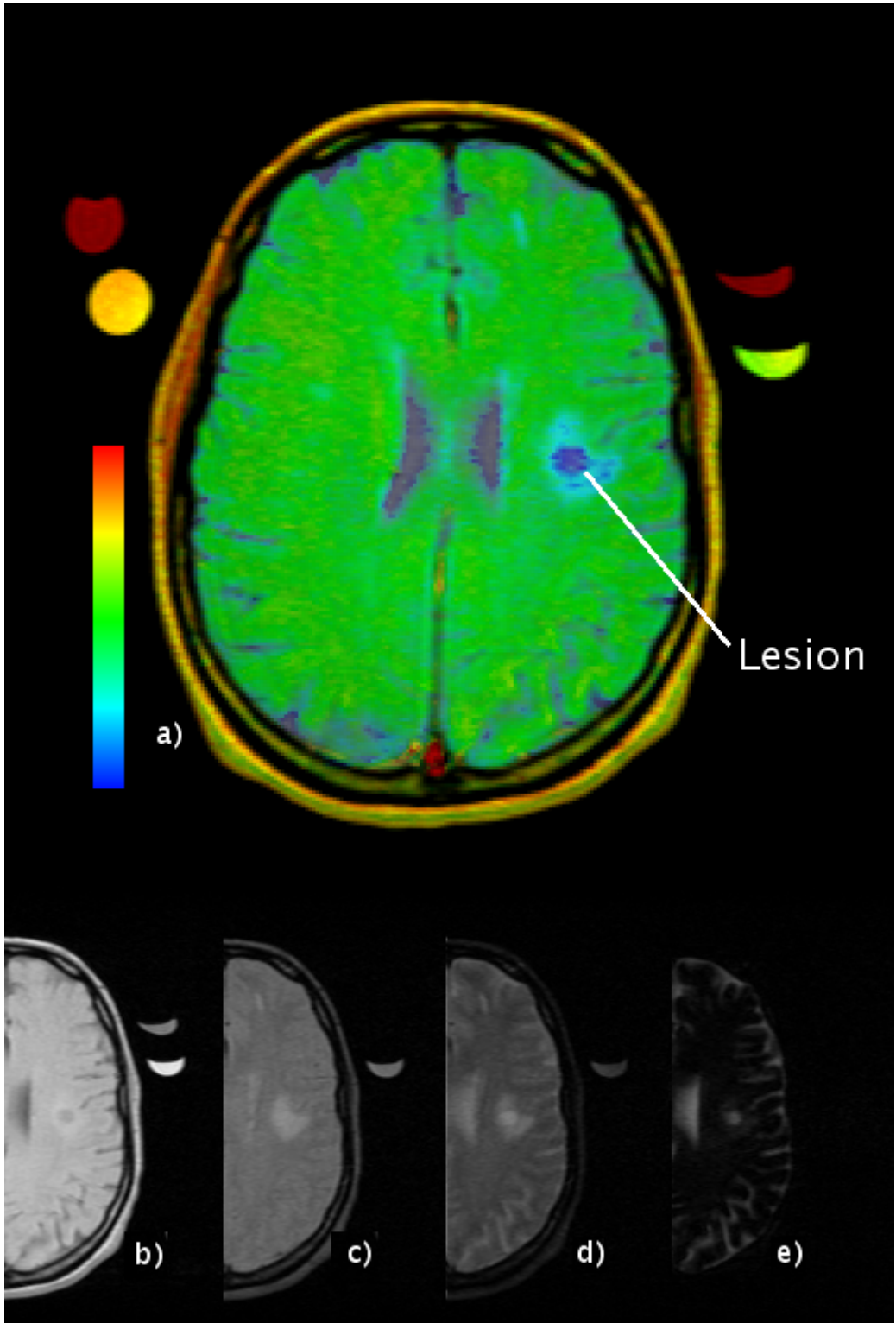

**Fig 2**

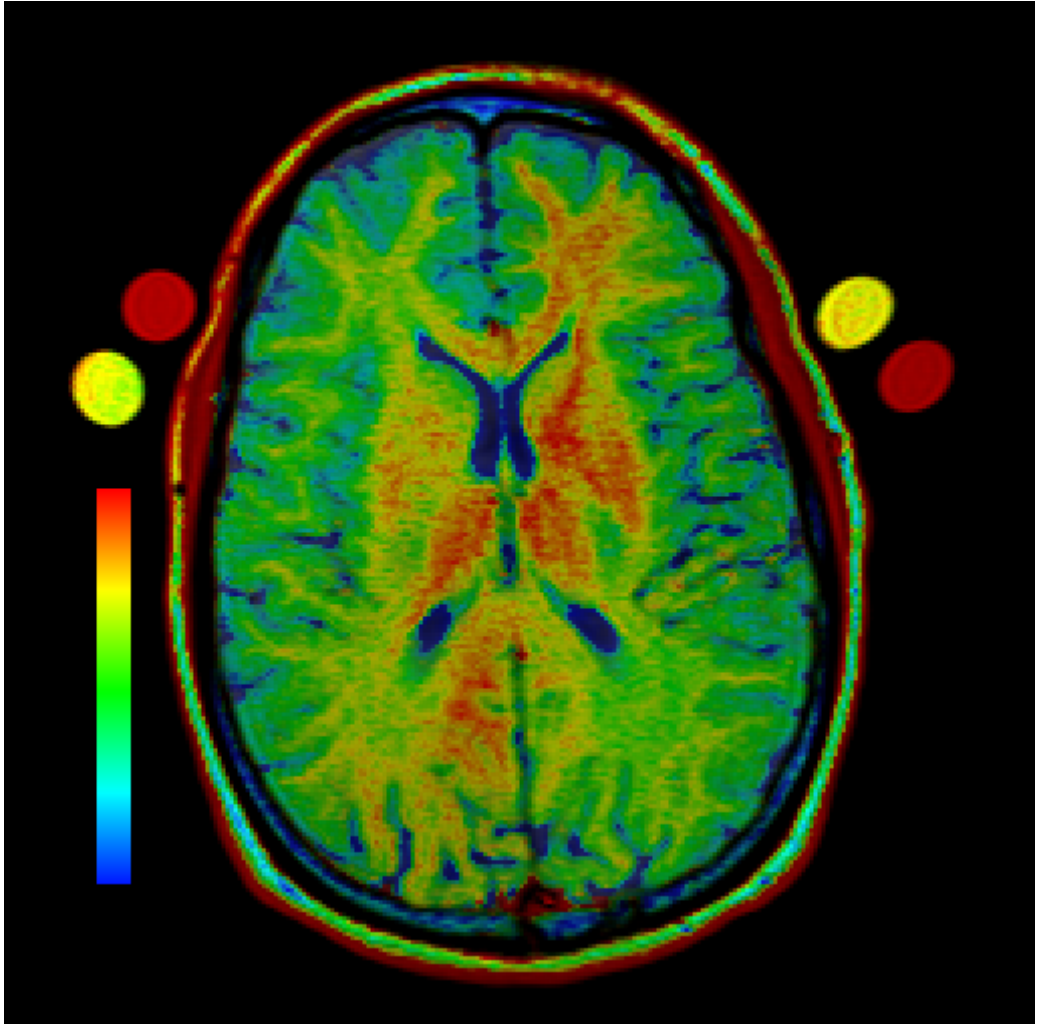

**Fig 3**

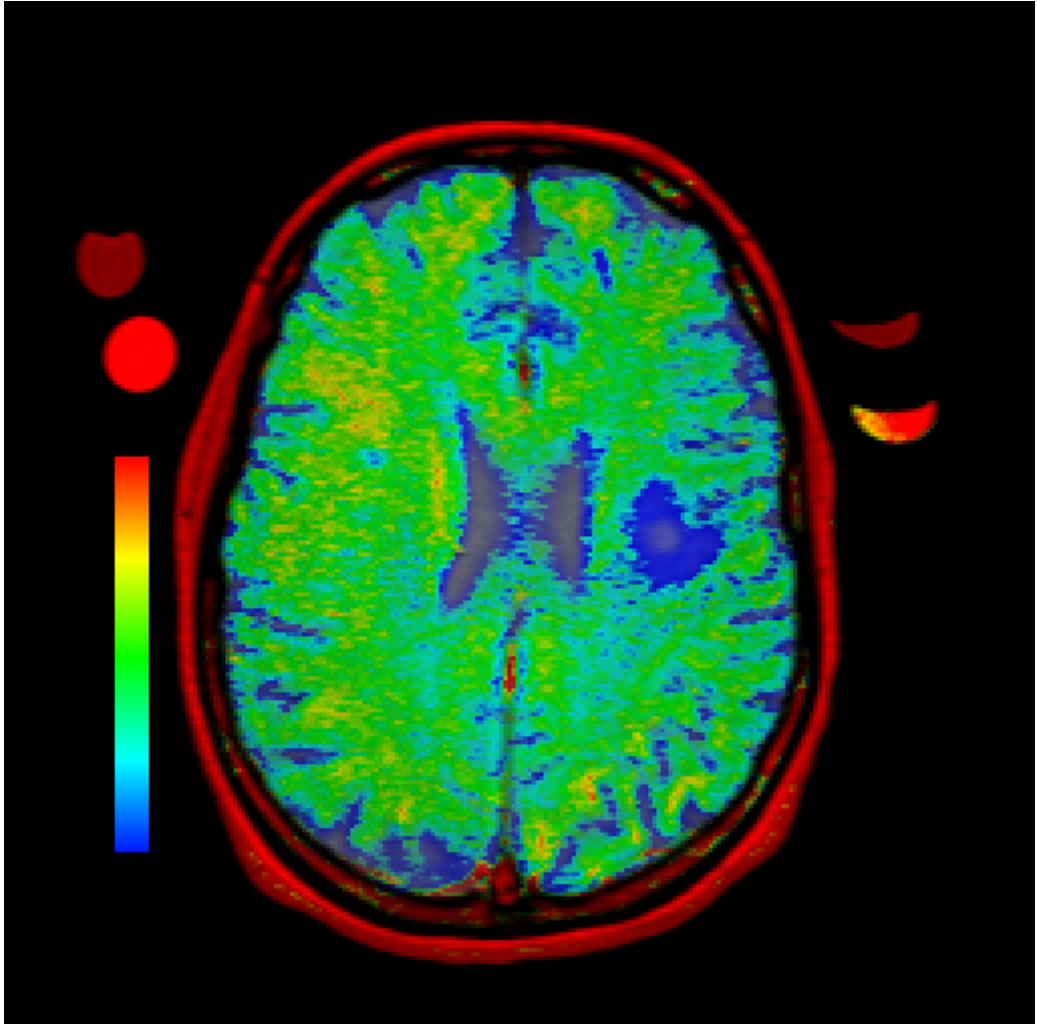

**Fig. 4**